# Computations of Nonlinear Propagation of Sound Emitted from High Speed Mixing Layers

J. Punekar[1], E.J. Avital[*,1] and R.E. Musafir[2]

[1]*School of Engineering and Materials, Queen Mary, University of London, Mile End Road, London E1 4NS, UK*

[2]*Universidade Federal do Rio de Janeiro, CP68503, Rio de Janeiro, 21941-972, Brazil*

**Abstract:** Non-linear sound propagation is investigated computationally by simulating compressible time-developing mixing layers using the Large Eddy Simulation (LES) approach and solving the viscous Burgers Equation. The mixing layers are of convective Mach numbers of 0.4, 0.8 and 1.2. The LES results agree qualitatively with known flow behavior. Mach waves are observed in the near sound field of the supersonic mixing layer computed by the LES. These waves show steepening typical to non-linear propagation. Further calculations using the Burgers equation support this finding, where the initial wave slope has a role in kicking them. No visible non-linear propagation effects were found for the subsonic mixing layers. The effects of geometrical spreading and viscosity are also considered.

## 1. INTRODUCTION

Commonly, linear propagation is assumed for sound emitted by free shear flows such as jets. This assumption is suggested in the Lighthill acoustic analogy [1] and is also employed in approaches which account explicitly for mean flow effects on the sound emission, e.g. [2], having been used successfully to predict sound generated by subsonic and supersonic jets [2-4]. By these approaches the sound is generated by the deformation of the flow eddies as they convect along the jet. As the jet speed increases the emitted sound amplitude also increases as expressed by Lighthill's $U^8$ law for the emitted acoustic power output from subsonic jets. However, the latter still remains a tiny fraction of the hydrodynamic power embedded in the jet, thus justifying as it seems the linear acoustics assumption. The increase in the sound amplitude is more profound in the downstream side of the jet as exhibited e.g. by the Lighthill's Doppler factor $(1 - M_S \cos\theta)^{-5}$ for the acoustic intensity, where $M_S$ is the source Mach number and $\theta$ is the spherical downstream angle.

At supersonic speed this form of the Doppler factor breaks down at the Mach angle $\theta = \cos^{-1}(1/M_S)$. Intensive sound waves called Mach waves propagate at that direction [1]. They are caused by a change in the sound generation mechanism where the mere supersonic convection of the eddies generate the sound. A correction to the Lighthill's Doppler factor making it finite in the Mach wave direction while assuming linear acoustics is possible by taking into account the finite life span of the sound-generating flow eddies [5]. Another approach taking into account explicitly the high speed mean flow effects was given by Goldstein &

Leib [2], where attention was given to the peak of the supersonically generated noise. However, Goldstein & Leib [2] were careful to limit their computations to jets that were not dominated by Mach waves emission, arguing the latter could be affected by non-linear propagation effects, which were not accounted in their analysis. These non-linear (NL) effects can appear because of the change in the mechanism of the sound generation, causing the Mach waves to have high amplitude as relative to the subsonically-generated sound and possibly to have also a higher wave steepness.

Lighthill [1] used the non-linear inviscid planar Burgers equation to show that saw-tooth waves can damp as $1/r$ because of NL propagation effects, where $r$ is the propagation distance. However the non-linear process of "bunching" where waves merge can moderate that damping. It was postulated that this process might occur in the near sound field where Mach waves acted as plane waves. Crighton [6] argued for the importance of the Sound Pressure Level (SPL) to kick NL effects and used the Burgers equation to derive an asymptotic estimate for the decay of the frequency spectrum at the high frequency range. The planar Burgers equation was also used by Punekar *et al.* [7] to study the statistical evolution of random waves as may be produced by supersonic jets. Freund *et al.* [8] used data achieved from Direct Numerical Simulation (DNS) of a supersonic perfectly expanded jet and the 1D weak shock-wave theory to study NL propagation. They noticed NL effects in terms of amplitude reduction and steepening of the sound wave. Nevertheless, it was concluded that linear acoustics could still capture the main features of the sound propagation and a linear wave equation was further used.

Brouwer [9] used the Burgers equation to analyze NL propagation for jet noise emitted by the F18 jet fighter engine. The initial time signal that was fed into the Burgers equation was constructed by producing random waves and then adjusting their frequency spectra to published spectra from F18 engine tests. It was suggested to correlate NL propagation effects with the Sound Pressure Level (SPL) and the peak frequency in the near sound field. NL effects were shown to reduce the SPL up to 6dB when the starting point for the Burgers equation calculation was about 10 m away from the jet. As the starting point for the Burgers calculations was taken further away from the jet the NL effects diminished.

*Address correspondence to this author at the School of Engineering and Materials, Queen Mary, University of London, Mile End Road, London E1 4NS, UK; Tel: +44(0)20 7882 3616; Fax: +44(0)20 7882 5532; E-mail: e.avital@qmul.ac.uk

Gee *et al.* [10] analyzed ground F22 jet fighter engine tests using the Burgers equation as well. Unlike Brouwer [9] they found no NL effects when the engine operated without an aft burner. They reasoned it by the lower level of the F22 SPL as compared to the F18. Thus when the aft burner of the F22 engine was used, which raised the far field SPL by about 50 dB, noticeable NL effects were found in the high frequency range. Saxena *et al.* [11] developed a frequency domain solver for the Burgers equation taking into account atmospheric relaxation processes and ground reflections. They applied the solver for jet noise obtained experimentally from an anechoic chamber and F18 engine ground tests. NL effects in the high frequency range were noticed for heated jets with exit Mach number $M_J = 0.9$ & $1.9$ at about 150 diameters away from the jet. The F18 noise analysis showed a reduction in the peak amplitude and an increase in the high frequency range because of NL effects as was also found by Brouwer [9].

In this study we will look at NL propagation for noise emitted by the mixing region of a free jet with the aim of identifying Mach number dependence for the onset of non-linear processes. Other noise components emitted by non-perfectly supersonic jets such as broadband shock-cell and screech noise will be discussed elsewhere. Lighthill [1] postulated that NL effects can be important for supersonic sound sources because of the generation of the Mach waves. On the other hand, Crighton [6] and Gee *et al.* [10] put more emphasis on the SPL of the near field as the trigger causing NL propagation. Clearly Mach waves are associated with an increased SPL, but the question whether they are necessary and sufficient to trigger NL propagation remains open. Here the following roles in triggering NL propagation will be computationally studied; the Mach number of the sound source $M_S$ and the initial sound wave form including amplitude and steepness.

For the purpose of our investigation compressible time-developing planar mixing layers will be simulated. The density fluctuations recorded in the near sound field of the mixing layer will be propagated using the Burgers equation to examine NL propagation effects. The time-developing mixing layer occurs when periodic streamwise boundary conditions are assumed causing the layer to develop in time rather in the streamwise direction in terms of transition to turbulence. This kind of flow development requires much less computational effort than the spatially evolving mixing layer, although the latter resembles much better what is seen in experiments. Nevertheless, Avital *et al.* [12, 13] showed that the near sound field of supersonic time-developing mixing layers resembles that of the corresponding spatially-evolving mixing layers. This correspondence was used to analyze DNS results of supersonic time-developing mixing layers and to predict the behavior of supersonic jet mixing noise. Although the similarity breaks down for subsonic mixing layers, Fortuné *et al.* [14] used successfully DNS results of low speed time-developing mixing layers to reproduce known aeroacoustic results and examine the effect of the Mach number in the subsonic range. Hence the time-developing mixing layer will be used as the model shear flow in this study. The mathematical and computational formulations of the simulations and acoustic analysis are presented in the next section. This will be followed by analysis of results obtained for cold mixing layers of various Mach numbers.

## 2. MATHEMATICAL AND NUMERICAL FORMU-LATION

### 2.1. Compressible Mixing Layers Simulations

Subsonic to supersonic planar mixing layers were simulated using the Large Eddy Simulation (LES) approach. In this approach the flow large scale structures are simulated while the small scale structures are modelled. LES has been shown to be capable of producing the flow large scale structures accurately [15], which is of interest in this study. The Favre filtered compressible Navier-Stokes (NS) equations were simulated in rectangular co-ordinates:

$$\frac{\partial \rho}{\partial t} + \frac{\partial (\rho u_i)}{\partial x_i} = 0 , \qquad (1)$$

$$\frac{\partial (\rho u_i)}{\partial t} + \frac{\partial (\rho u_i u_j)}{\partial x_j} = -\frac{\partial p}{\partial x_i} + \frac{\partial \sigma_{ij}}{\partial x_j} - \frac{\partial (\rho \tau_{ij})}{\partial x_j} , \qquad (2)$$

$$\frac{\partial E}{\partial t} + \frac{\partial \left[ u_j (E + p) \right]}{\partial x_j} + \frac{\partial q}{\partial x_i} = \frac{\partial (\sigma_{ij} u_j)}{\partial x_j} , \qquad (3)$$

where $\rho$ is the density, $u_i$ is the velocity component in the $i$ direction, $p$ is the pressure and $E$ is the total energy. The conventional over bars and tildes noting the flow properties as spatially filtered were omitted from Equs. (1) to (3) for simplicity. $\sigma_{ij}$ is the molecular viscosity calculated by assuming Newtonian fluid and using Sutherland's formula connecting the molecular viscosity coefficient with the temperature variation [15, 16]. $\rho \tau_{ij}$ is the LES sub-grid scale tensor representing the effect of the small scale flow structures on the simulated large scale structures, which was calculated using the dynamic Smagorinsky model [15]. $q$ is the heat flux vector calculated using Fourier's law. The sub-grid terms in the energy equation (3) were omitted because they have a very small effect on the flow momentum terms [15], which are the terms responsible for the mixing noise emitted by cold layers as investigated in this study. The NS equations (1) to (3) are supplemented by assuming perfect gas and using the equation of state.

The mixing layer was constructed by specifying two opposing streams with the same speed and ambient properties of density and temperature. The convective Mach number of the mixing layer $M_C$ is defined as

$$M_C = \frac{U_1 - U_2}{c_{01} + c_{02}}, \qquad (4)$$

where $U_1$ and $U_2$ are the upper and lower streams ambient velocity respectively, and $c_{01}$ and $c_{02}$ are the upper and lower streams ambient speed of sound respectively [15, 16]. Thus in our case $M_C$ is simply the ambient Mach number of the streams. The velocity, density and temperature fields were all normalized by the corresponding ambient values before solving the NS equations. The spatial dimensions were normalized so that one unit length was equal to twice the initial mean momentum thickness $\delta$, where the initial mean streamwise velocity component was prescribed using a hyperbolic-tangent profile [14-16]. The other initial mean velocity components were taken as zero and the initial mean density and temperature fields were taken as uniform having the ambient values. Thus the mixing layers started from an isothermal condition.

Periodic boundary conditions were used in the streamwise and spanwise directions, thus the mean momentum thickness $\delta$ was defined as

$$\delta \equiv \frac{1}{4} \int_{-L_y/2}^{L_y/2} <\rho> \left(1 - \frac{<\rho u>}{<\rho>}\right)\left(1 + \frac{<\rho u>}{<\rho>}\right) dy, \qquad (5)$$

where $L_y$ is the stream-normal length of the computational box and the operator $<\ >$ denotes spatial averaging in the streamwise and spanwise directions. Non-reflecting boundary conditions were used at the stream-normal direction, i.e. at $y = \pm L_y/2$ [17]. Uniform grid was taken at the streamwise and spanwise directions. A stretched grid in the stream-normal direction was taken as in Refs. [12, 15], giving a ratio of about 8 to 1 between the grid spacing at the edges of the computational box to that at its centre.

Collocated grid and finite difference schemes were used to discretize the NS equations (1) to (3). Because of the possibility of shocklets inside the mixing layer [15], the Steger-Warming flux splitting scheme was used for the convection terms [18] along with the 5th order Weighted Essentially Non-Oscillatory (WENO) differentiation scheme of Jiang & Shu [19]. A fourth order central scheme was used for the diffusion terms. The NS equations were marched in time using a compact 3rd order Runge-Kutta scheme. To start the transition to turbulence, initial 3D disturbances of oblique waves and imposed random noise were used. Maximum overall disturbance amplitude of 0.1 was used at the centre of the mixing layer, while that amplitude decayed exponentially towards the streamwise edges of the computational domain, as in Fortuné *et al.* [14].

## 2.2. Burgers Equation Solution

The density fluctuations denoted as $\rho'$ were recorded near the top of the LES computational domain and fed as the boundary condition into a generalized viscous non-linear Burger equation to investigate NL propagation. The Burgers equation was commonly used as a tool to investigate non-linear propagation as described in the Introduction section. That equation was taken as

$$\frac{\partial \rho'}{\partial t} + \frac{m\rho'}{r} + \frac{\partial}{\partial r}\left[c_0\left(1 + \frac{\beta\rho'}{2\rho_0}\right)\rho'\right] = \frac{\upsilon}{2}\left(\frac{4}{3} + \frac{\mu_B}{\mu} + \frac{\gamma-1}{\text{Pr}}\right)\frac{\partial^2 \rho'}{\partial r^2}, \qquad (6)$$

where $r$ is the propagation distance and $m = (0, 0.5, 1)$ for planar, cylindrical and spherical spreading, respectively [20]. The non-linearity and diffusion terms are expressed in Eq. (6) using spatial derivatives in order to fit for the employed numerical scheme. The more commonly version used in acoustics expresses these terms using temporal derivatives. Both versions are equivalent when considering the approximations that led to the derivation and applicability of the Burgers equation [20].

The non-linearity coefficient $\beta$ was taken as $(\gamma + 1)/2$ as appropriate for perfect gas [20], with $\gamma = 1.4$ for air. If $\beta = 0$ then Equation (6) will become fully linear and will be denoted as the linearised Burgers Equation. Usually, the linear propagation term $\partial(c_0\rho')/\partial r$ is removed from Eq. (6) by replacing the time co-ordinate $t$ with the retarded time co-ordinate $\tau \equiv t - r/c_0$ [20]. It was kept in this study in order to ease the comparison between linear and non-linear propagation as it will be shown later. $\upsilon$ and $\mu$ are the shear kinematic and kinetic viscosity coefficients respectively, and $\mu_B$ represents the bulk viscosity. $\mu_B/\mu$ is taken as 0.6 for air, approximating the effect of the relaxation process as purely dissipative, which is appropriate for the low frequency range [20]. The Prandtl number $Pr$ was taken as 0.7.

A uniform grid and finite difference schemes were used to discretize Eq. (6). The LES density fluctuations were fed as the boundary condition at $r = 0$ as stated earlier and an inviscid non-reflecting boundary condition was used on the other edge of the grid at $r = R$. $\rho'$ was taken as zero at $t = 0$. The 5th order WENO scheme was used to approximate the propagation terms and a 4th order central scheme was used for the diffusion term. The equation was marched in time using a 2nd order compact Runge-Kutta scheme.

To test the Burgers equation solver a sine wave was fed as the boundary condition at $r = 0$. This should yield the Fubini solution in the near field for the inviscid planar Burgers equation [20, 21], expressed as

$$\rho' = 2\rho_{amp}\sum_{n=1}^{\infty}\frac{J_n(nr/l)}{nr/l}\sin\left[n\left(\omega t - kx\right)\right], \qquad (7)$$

where

$$l = \frac{\rho_0 c_0}{\omega\beta\rho_{amp}}, \quad k = \frac{\omega}{c_0}, \qquad (8)$$

and $J_n$ is the Bessel function of the first kind. The above solution is valid for the pre-shock region of $r < l$ and $\rho'$ becomes $\sin(\omega t)$ at $r = 0$. A comparison between the Fubini solution and the numerical solution of the Burgers equation (Eq. 6) for $m = \upsilon = 0$ is shown in Fig (**2**). $(\rho_0, c_0)$ were taken as (1, 10) respectively and $l$ as 100. Thus the non-linear propagation term $c_0\beta\rho'/(2\rho_0)$ in Eq. (6) was less than 1% of the linear term $c_0\rho'$. Nevertheless, the numerical solution successfully produced the accumulative NL propagation effects of wave steepening while preserving the wave amplitude and achieving a very good agreement with the Fubini solution. The grid resolution used for the numerical solution shown in Fig. (**2**) was about 30 points per wave length. Increasing the resolution to about 40 points per wave length removed completely the small differences in the wave troughs between the two solutions that can be seen in Fig. (**2b**). This type of resolution may be considered to be too high for multi-dimensional problems but is more than adequate for the current 1D problem of Eq. (6).

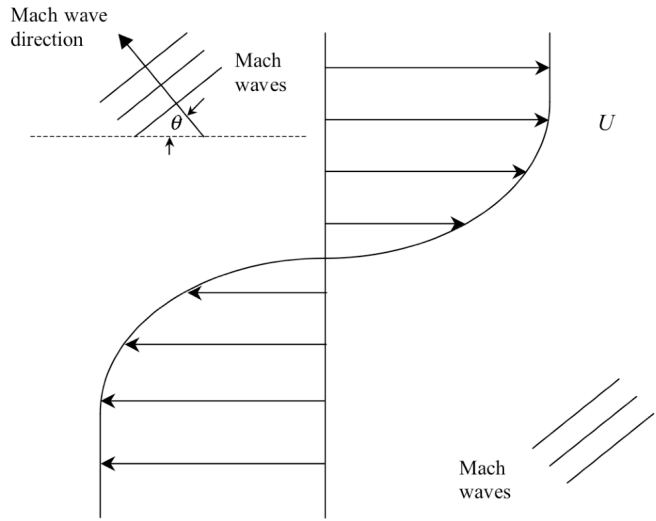

**Fig. (1).** Schematic description of the mixing layer and emitted Mach waves.

## 3. RESULTS

A mixing layer with a *tanh* streamwise velocity profile was initialized by an oblique disturbance with a random phase to enhance transition to turbulence. The Reynolds number based on the initial momentum thickness was set to 500 and the convective Mach numbers values of 0.4, 0.8 and 1.2 were considered. The computational domain size was taken as $(Lx, Ly, Lz) = (20, 40, 20), (25, 40, 25)$ and $(30, 40, 30)$ for $M_C = 0.4, 0.8$ and 1.2, respectively, where one spatial unit was equivalent to twice the size of the initial momentum thickness as used by Vreman *et al.* [16]. The computational grid size was taken as of (61, 241, 61), (88, 281, 88) and (101, 321, 101) points, respectively, so that the number of grid points increases with $M_C$. This kind of resolution scales favourably with the grid resolution in Vreman [15] when considering Reynolds number scaling.

The time evolution of the mean momentum thickness $\delta$ is shown in Fig. (**3**) for the various simulated mixing layers. All mixing layers initially show a mild increase in the momentum thickness as ambient fluid is entrained into the mixing layer. This is followed by a rapid increase in the momentum thickness marking the flow transition to turbulence. As expected, increasing the mixing layer's Mach number delays the transition to turbulence and reduces the rate of the momentum thickness growth [15, 16]. A fully developed mixing layer will exhibit a linear growth of the momentum thickness with time. All mixing layers show a nearly linear growth at later time stages in Fig. (**3**), although none shows exactly a linear growth because of the finite size of the computational domain. The growth rates were calculated as (0.07, 0.025, 0.02) for $M_C = (0.4, 0.8, 1.2)$, while Vreman *et al.*'s [16] DNS-based model produced the growth rates of (0.065, 0.035, 0.025) and a linear-stability based model produced the growth rates of (0.06, 0.033, 0.02), respectively. Thus a fair to good agreement has been achieved where the highest difference is for the $M_C = 0.8$ case. However, as it will be shown next, all mixing layers achieved flow development close to a self-similar state.

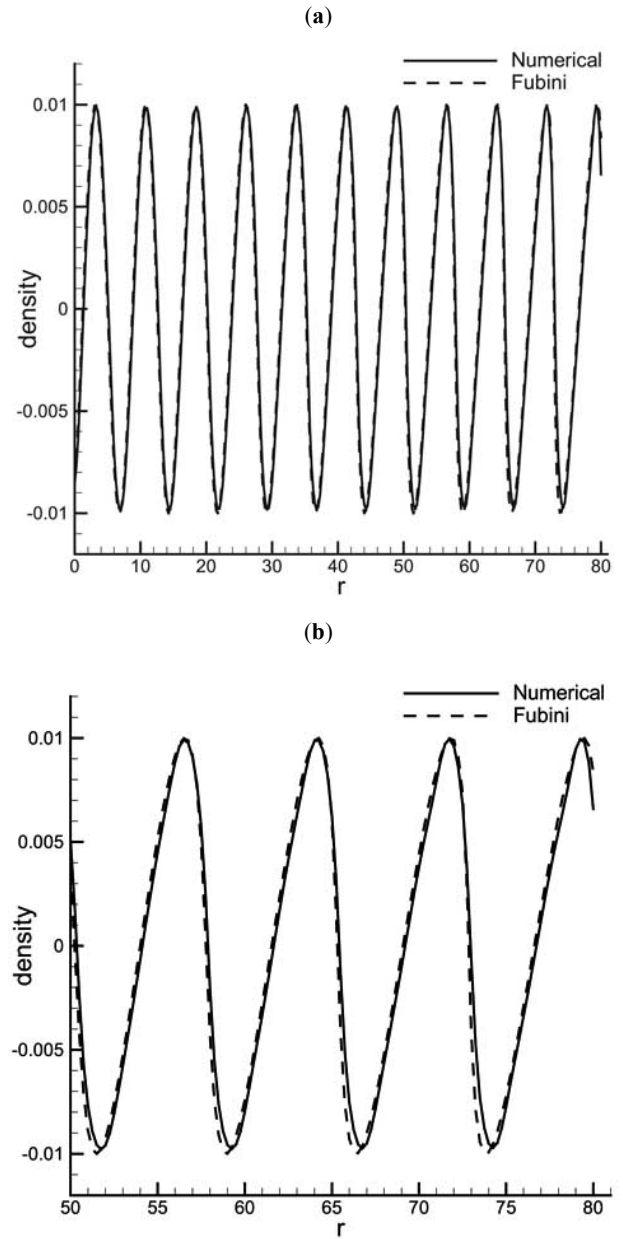

**Fig. (2).** Comparison between the numerical solution for the plane inviscid Burgers Equation (6) and the theoretical Fubini solution (7) for the instantaneous density fluctuations of a monochromic wave starting as a sinusoidal wave at $r = 0$.

The mean streamwise velocity profile is shown in Fig. (**4**) for the supersonic mixing layer of $M_C = 1.2$. An empirical curve of a fully developed mixing layer [22] is also shown, demonstrating that up to late time stages the simulation managed to preserve a velocity profile close to the empirical one. Similar behaviour was found with the other mixing layers. The preservation of an almost fully developed mixing layer is further illustrated in Fig. (**5**) showing the stream-normal variation of the mean turbulence intensity defined as $\langle \rho v_i v_i \rangle^{0.5}$, where $v_i = u_i - \langle \rho u_i \rangle / \langle \rho \rangle$ is the velocity fluctuation. The profiles coincide well for all mixing layers when the stream-normal direction is normalized by the mean momentum thickness. The levels of the turbulence intensity also scale favorably with the levels reported by Vreman [15], demonstrating that a self-similar state was reached.

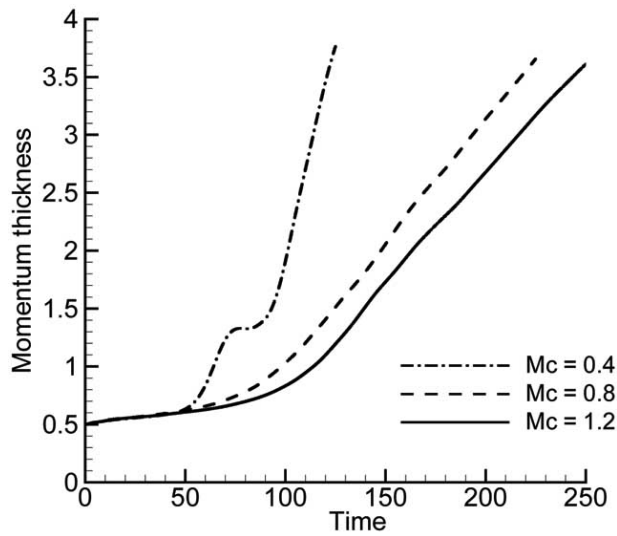

**Fig. (3).** The time evolution of the mean momentum thickness for the simulated mixing layers.

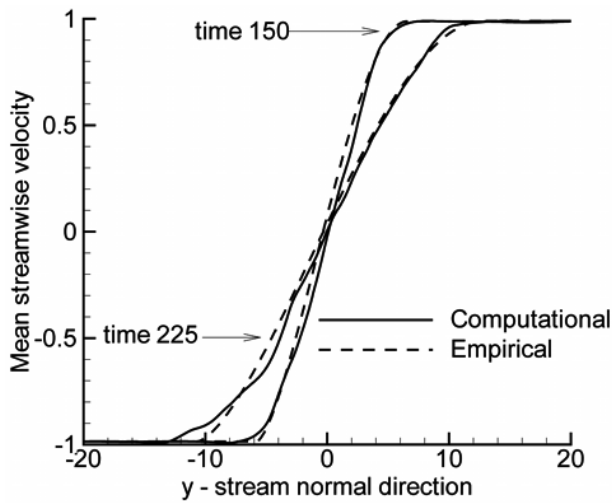

**Fig. (4).** Mean streamwise velocity profiles for the supersonic mixing layer of $M_C = 1.2$, where the empirical curve was taken from Abramovich [22].

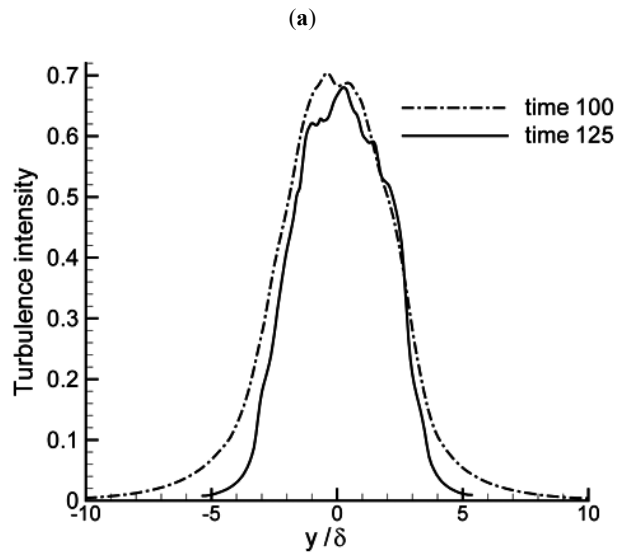

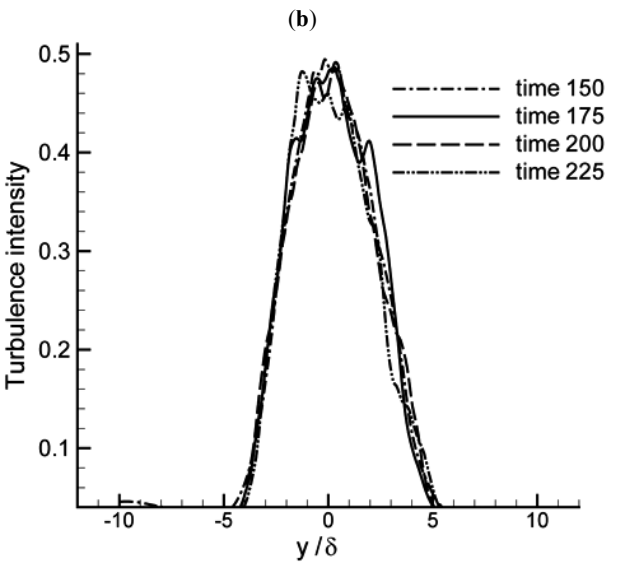

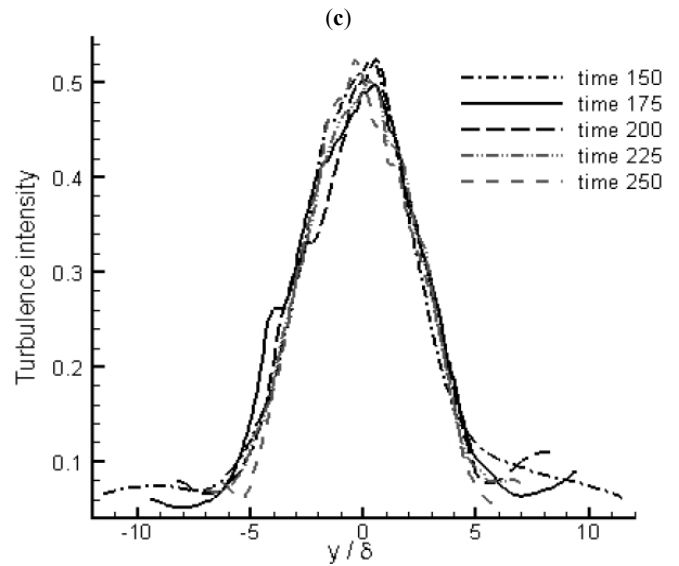

**Fig. (5).** The stream-normal variation of the mean turbulent kinetic energy for the mixing layers of (**a**) $M_C = 0.4$, (**b**) $M_C = 0.8$ and (**c**) $M_C = 1.2$.

Typical instantaneous density contours are shown in Figs. (**6**, **7**) for the various mixing layers in the mid-plane of the computational box in the spanwise direction. It can be seen that as $M_C$ increases higher density fluctuations appear outside the mixing layer as expected. However, while the low speed mixing layer of $M_C$=0.4 shows no inclined wavy structures outside the shear zone, the contours of the supersonic mixing layer of $M_C$=1.2 show such structures as indicated in Fig. (**7**). Furthermore these structures which we identified as Mach waves can show steepening in the propagation direction as seen in Fig. (**7b**). This is a sign of a non-linear propagation effect. The propagation angle $\theta$ (see Fig. **1**) is about $45^0$, which gives a Mach number $M_S$ of about 1.4 for the source of the wave, when taking $M_S = 1/\cos\theta$. Considering that the lower stream moves at a Mach number of 2.4 (twice of $M_C$) as relative to the upper stream, $M_S = 1.4$ gives a ratio of about 0.6 when normalized by 2.4, which is well inside the range of 0.5 to 0.7 for the expected speed of sound sources in a jet [1-3]. On the other hand the high subsonic mixing layer of $M_C = 0.8$ shows a weak inclined structure in the lower stream and a vertical one at the upper stream as indicated in Fig. (**6b**). Since most of the sound emission in this mixing layer is generated by subsonic flow there can still be some cancellation between the forward and aft sound emissions caused by the periodic streamwise boundary conditions [12]. This will result in wave structures tending to be vertical. Nevertheless some of the sound source still convects supersonically as relative to the upper or lower stream, causing a mild inclined wave structure as seen in Fig. (**6b**).

The density fluctuations recorded at several points in the upper stream of the mixing layers were propagated using the Burgers equation (6). Typical time histories of the sound wave propagation obtained for the inviscid planar case are shown in Figs. (**8**, **9**). Results from the linearised form ($\beta = 0$) are also included in Fig. (**9**). Both subsonic mixing layers of $M_C = 0.4$ and 0.8 show linear propagation characterized by parallel trough and crests lines although the Burgers equation (6) that was considered accounted for non-linear propagation. On the other hand the sound propagation for the supersonic mixing layer shows non-linear characteristics of crests moving faster then troughs, leading to merger and cancellation as seen for the trough-crest pairs starting at about times 30 and 180. This is particular evident when comparing with the solution of the linearised Burgers equation ($\beta = 0$) shown in Fig. (**9b**). Further evidence to the higher speeds of the crests is seen for the crests starting around time 150.

To assess the effect of geometrical spreading, the propagation calculations were repeated for the inviscid cylindrical Burgers equation. The results are shown in Fig. (**10**). The cylindrical spreading can be observed as a dominant feature affecting the propagation pattern by reducing the wave amplitude to a similar level whether linear or non-linear propagation is assumed. However, non-linear effects are still evident as demonstrated by the mergers of the trough-crest pairs indicated in Fig. (**10a**). To better assess the effects of non-linearity and geometrical spreading the time histories of the density fluctuations were plotted in Fig. (**11**) for the propagation distance 40, i.e. eighty times the initial momentum thickness of the mixing layer. In the case of planar propagation, crests or troughs with an amplitude as little as 0.05 already show non-linear behaviour by moving faster or slower, respectively. More striking is the damping of the trough-crest pair around time 230. Although the pair had amplitudes originally similar to those of other crests and troughs, the non-linear effect is much more evident.

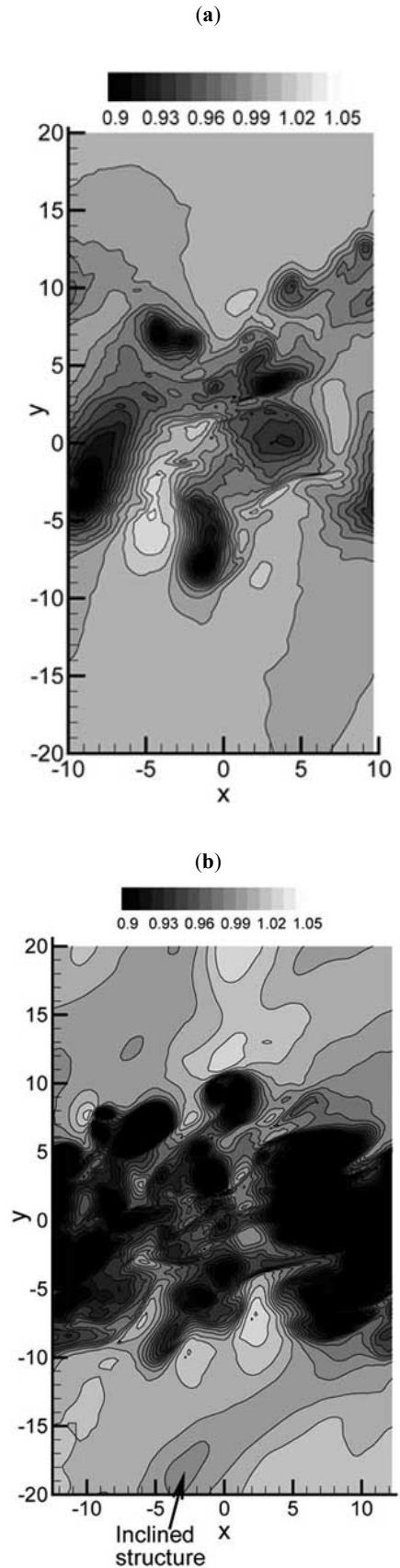

**Fig. (6).** Typical instantaneous density contours at the mid spanwise plane ($z = 0$) for the mixing layers of (**a**) $M_C = 0.4$ and time 120; (**b**) $M_C$=0.8 and time 200.

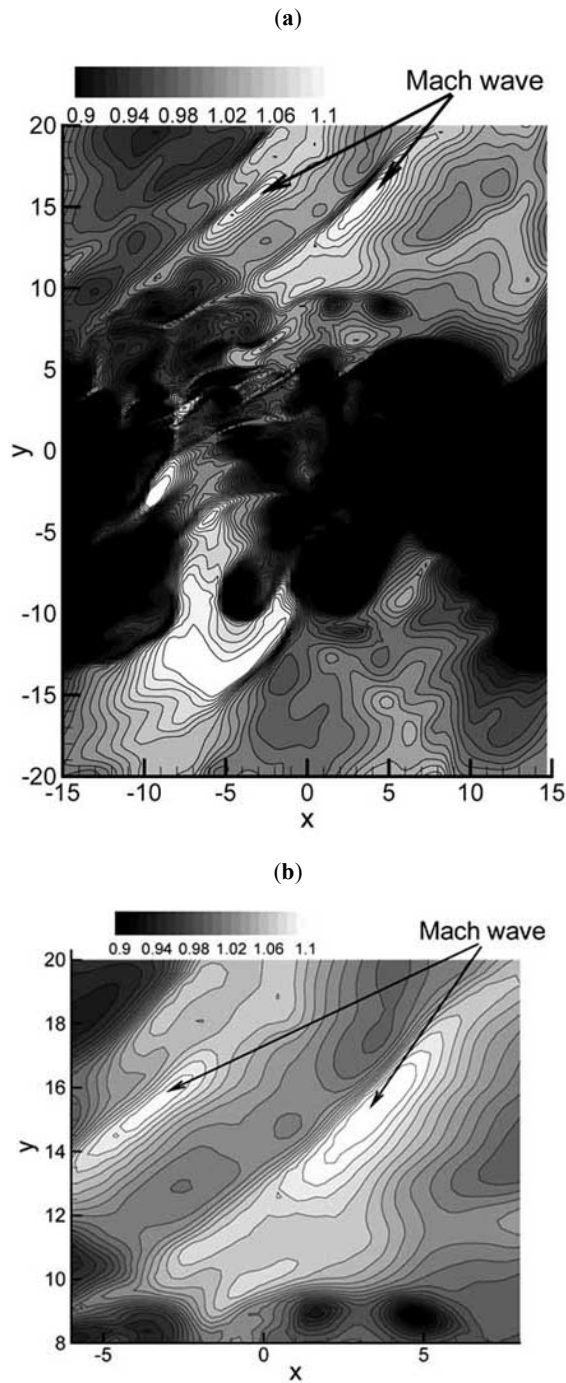

**Fig. (7).** Typical instantaneous contour levels at the mid spanwise plane ($z = 0$) for the supersonic mixing layer of $M_C = 1.2$ and time 225, and for (**a**) the entire mid-plane, and (**b**) a zoomed view on Mach waves.

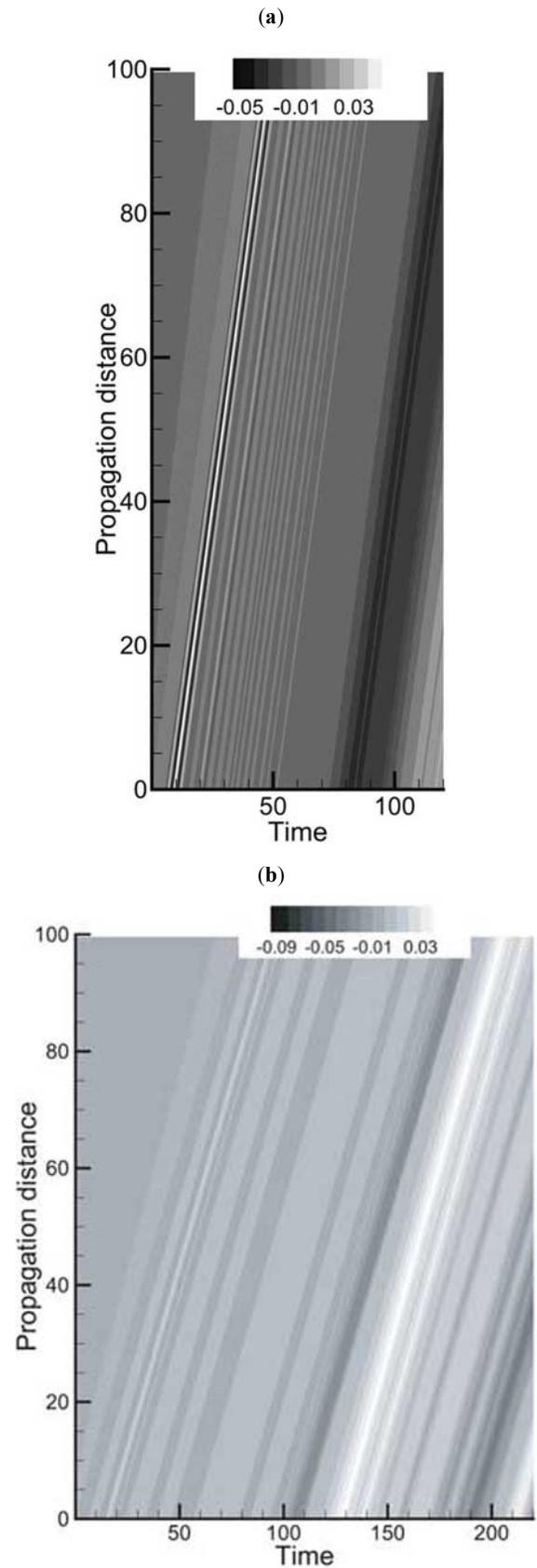

**Fig. (8).** The propagation of the density fluctuations as calculated using the inviscid plane non-linear Burgers Equation (6) for the mixing layers of (**a**) $M_C = 0.4$ and (**b**) $M_C = 0.8$. The density fluctuations at the zero propagation distance were taken from the LES at the computational domain point of $(x, y, z) = (-2.5, 12, 0)$.

**(a)**

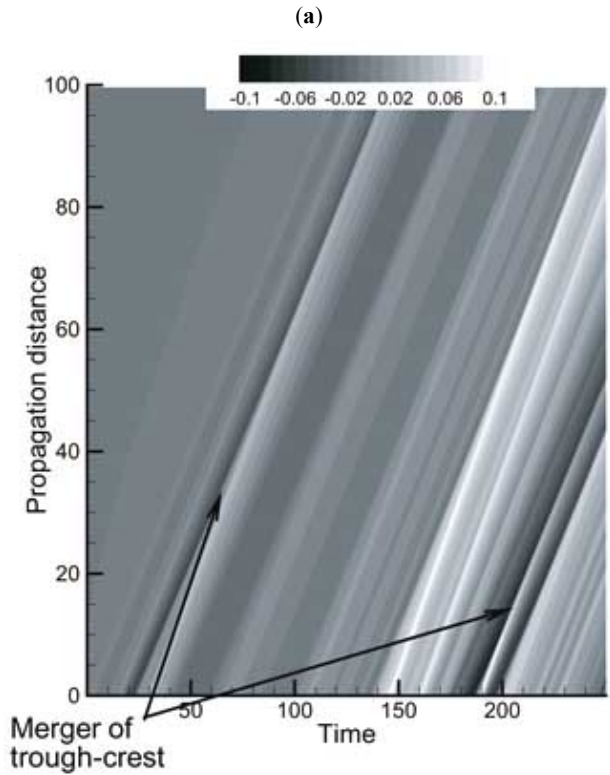

**(a)**

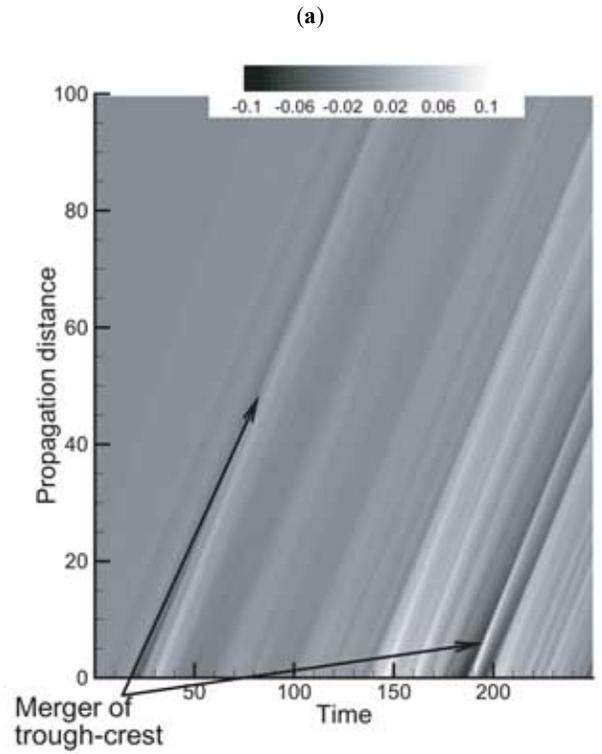

**(b)**

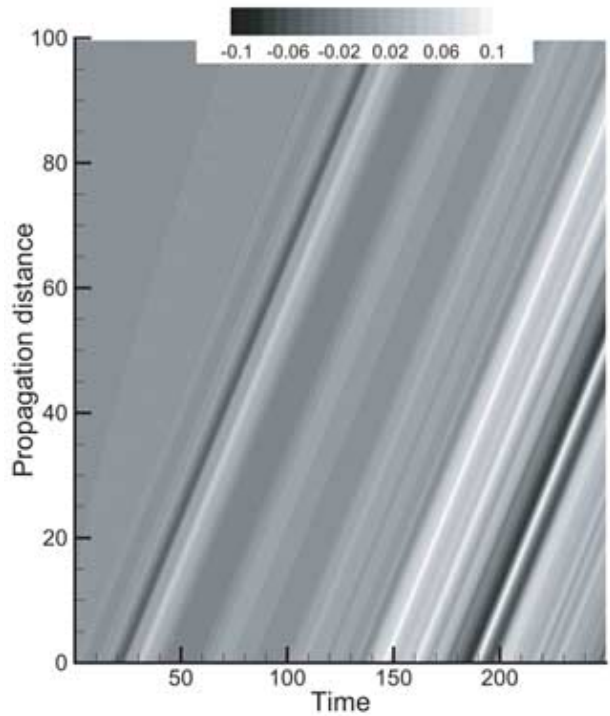

**(b)**

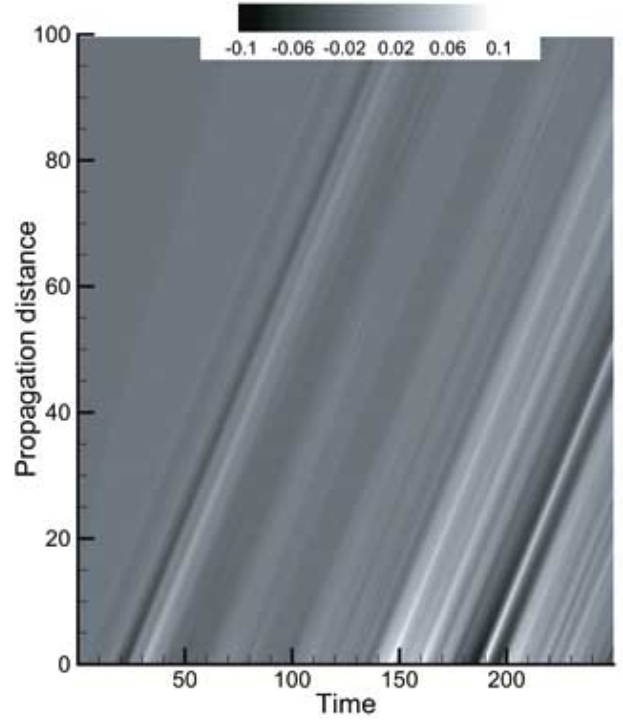

**Fig. (9).** The propagation of the density fluctuations as calculated using the inviscid plane (**a**) Burgers Equation (6) and (**b**) its linearised form with $\beta = 0$, for the mixing layer of $M_C = 1.2$. The rest of the conditions are as in Fig. (**8**).

**Fig. (10).** The propagation of the density fluctuations as calculated using the inviscid cylindrical (**a**) Burgers Equation (6) and (**b**) its linearised form with $\beta = 0$, for the mixing layer of $M_C = 1.2$. The rest of the conditions are as in Fig. (**8**).

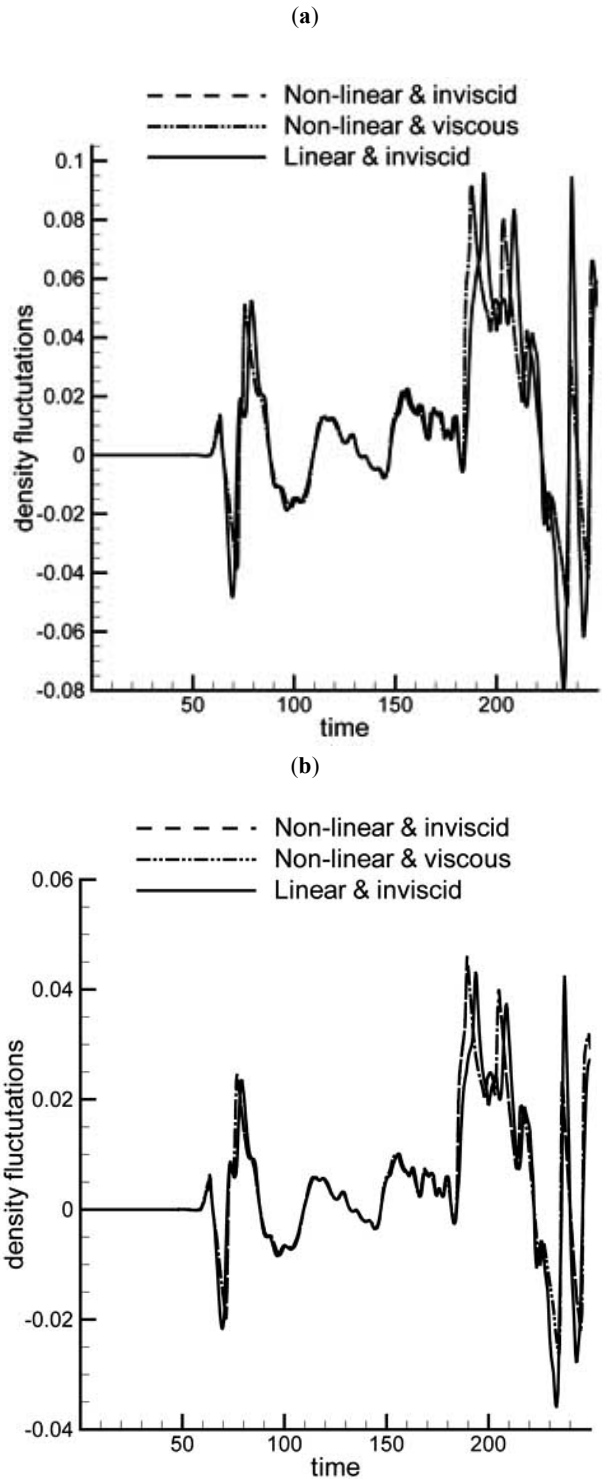

**Fig. (11).** Time history of the density fluctuations at a propagation distance of 40 assuming (**a**) planar propagation and (**b**) cylindrical propagation.

The non-linear effect is clearly overtaken by the damping caused by the cylindrical spreading seen in Fig. (**11b**), but still a significant non-linear effect is seen with the sharp trough-crest pair around time 230 as in planar propagation. This indicates that the sharp slope of the density fluctuation is at least as important as the amplitude in kicking non-linear propagation. Thus as it can be seen from both Fig. (**11a, b**), while the general pattern of the sound propagation can be predicted with some accuracy assuming linear propagation as suggested by Freund *et al.* [8], there are some events that can be completely miscalculated as the sharp trough-crest pair around time 230. On the other hand adding viscosity as modelled in Eq. (6) and using the Reynolds number of the simulated mixing layers had very little effect on the density fluctuations as can be seen in Fig. (**11**).

## 4. SUMMARY

Non-linear propagation of mixing noise was investigated by simulating low to high speed mixing layers using the Large Eddy Simulation (LES) approach and propagating the emitted sound using the Burgers equation. Time-developing cold mixing layers of convective Mach number $M_C = 0.4$, $0.8$ and $1.2$ were considered. Transition to turbulence and flow development close to self-similar state were observed. Generally good qualitative agreement was achieved with available results for the mixing layers' flow development. The Burgers equation was solved using a high order flux splitting scheme achieving a very good agreement with the Fubini analytical solution, pointing to the possibility of using such schemes for non-linear acoustic computations.

Instantaneous density fluctuations captured from the LES computations showed the generation of Mach waves by the supersonic mixing layer and steepening of the waves, pointing to non-linear propagation effects. No such wave formations were found for the subsonic mixing layers. Further propagation calculations using the Burgers equation showed non-linear propagation effects for the supersonic mixing layer, particularly when planar wave propagation was assumed. This result supports the hypothesis of Lighthill [1] of non-linear propagation affecting the near field region of Mach waves. It was seen that the initial steepening of the sound wave was as least as important as the wave amplitude in kicking non-linear effects. These findings showed that although the general pattern of the wave propagation can be predicted assuming linear acoustics, there are some steep peaks and troughs that will be significantly miscalculated. On the other hand viscosity effects were found to have little effect for propagation distances of up to hundred times of the initial momentum thickness of the mixing layer.

The subsonic or supersonic character of the sound source Mach number and the initial slope of the emitted sound wave were found to be crucial in determining non-linear propagation effects in low frequency mixing noise. Further planned studies include computations and performing accompanying experiments of spatially-evolving mixing layers. It is hoped that they will offer better mapping of the near sound field, allowing the identification of any transition from a planar wave form to a geometrically spreading wave and a better statistical analysis of non-linear propagation as it affects for example the frequency spectrum.

## ACKNOWLEDGEMENTS

The support given by the UK Daphne Jackson trust is kindly acknowledged as well as the support given by the Royal Society under its international joint project scheme.